\documentclass[12pt]{amsart}
\usepackage{amscd,amssymb}
\usepackage{epsfig}

\newtheorem{theorem}{Theorem}[section]

\theoremstyle{definition}

\theoremstyle{remark}

\newtheorem{example}[theorem]{Example}

\newcommand{\abs}[1]{\lvert#1\rvert}
\def\norm#1{\left\Vert#1\right\Vert}
\newcommand{\var}{\operatorname{var}}
\newcommand{\xv}{\mathbf{x}}

\newcommand{\yv}{\mathbf{y}}
\newcommand{\R}{\mathbb{R}}
\newcommand{\A}{\mathcal{A}}
\newcommand{\e}{\varepsilon}

\begin{document}

\title[Additive models in high dimensions]{Additive models in high dimensions}

\author{Markus Hegland}
\address{M.H.: CMA, School of Mathematical Sciences, Australian National
University, ACT 0200, Australia}
\email{Markus.Hegland@anu.edu.au}
\urladdr{http://datamining.anu.edu.au}
\author{Vladimir Pestov}
\address{V.P.: School of Mathematical and Computing Sciences,
Victoria University of Wellington, P.O. Box 600, Wellington, New Zealand}

\email{vova@mcs.vuw.ac.nz}
\urladdr{http://www.vuw.ac.nz/$^\sim$vova}

\address{{\it After July 1, 2002:} 
Department of Mathematics and Statistics,
University of Ottawa, Ottawa, Ontario, K1N 6N5, Canada}

\begin{abstract}
We discuss some aspects of approximating functions on high-dimensional
data sets with additive functions or ANOVA decompositions, that is,
sums of functions depending on fewer variables each. It is seen that
under appropriate smoothness conditions, the errors of the ANOVA 
decompositions are of order $O(n^{m/2})$ for
approximations using sums of functions of up to $m$ 
variables under some mild restrictions on the (possibly dependent)
predictor variables.
Several simulated examples illustrate this behaviour.
\end{abstract}


\maketitle

\section{Introduction}

In this paper we want to understand some aspects of
behaviour of additive predictive models \cite{HT}
for high-dimensional data sets.
More specifically, we will deal with the problem of approximating
a real-valued predictor function $f$ defined on a 
domain $\Omega$ and depending on
variables $x_1,x_2,\ldots, x_n,\ldots$ by additive functions
$f_{\mathrm{add}}$ in fewer (say $n$) variables and
of lower order of interaction, say $m$, having the form
\begin{eqnarray}
\label{eq:anova}
f_{\mathrm{add}}(x) &=& f_0 + f_1(x_1) +f_2(x_2) + \cdots + f_n(x_n) + \\
\nonumber
 & &
 f_{1,2}(x_1,x_2) + \cdots + f_{i_1,i_2}(x_{i_1},x_{i_2}) 
+ \cdots f_{n-1,n}(x_{n-1},x_n) + \cdots +
\\
\nonumber & &
f_{i_1,i_2,\cdots,i_m}(x_{i_1},x_{i_2},\cdots,x_{i_m}) + \cdots
+ f_{n-m+1,\cdots,n}(x_{n-m+1},\cdots, x_n),
\end{eqnarray}
where typically $m\ll n$. Our aim will be to obtain easily verifiable
upper bounds on the $L_2$-norm of the remainder $f-f_{\mathrm{add}}$, 
as well as to
try and understand their dependence on the dimension $n$ of the 
domain $\Omega$. In doing so, we will move away from the condition of
independence of the predictor variables $x_1,x_2,\ldots$, replacing
it with a milder restriction of the 
probability distribution being equivalent to the product of its marginals.
Our approximation with additive functions is optimal 
(with regard to the mean square error) for independent variables,
but not necessarily in the dependent case, where we obtain an
upper error bound. At the same time, we are not yet ready to offer
concrete algorithms for real very large datasets.

In one of its forms, the 
phenomenon of concentration of measure
\cite{Gr, M2, Ta}
says that every Lipschitz function on a sufficiently
high-dimensional domain $\Omega$ is 
well-approximated by a \emph{constant
function}, that is, an additive function of the
lowest possible order of interaction $m=0$. 
However, as one would expect, the limitations of this result are such as to
render it inapplicable in our situation: a reasonably 
good approximation requires the intrinsic dimension of a dataset to be
prohibitively high.

The most natural question is therefore, 
can one achieve a better approximation
in lower (mid to high) dimensions by merely
allowing additive functions of a higher interaction order $m>0$? 
Even here the answer turns out to be negative: there exist
functions for which
approximation by constants is the best possible among all additive functions
in the orders up to $m=n-1$. This result
makes it clear that the only way to achieve a
better approximation by additive functions is to impose further
restrictions on the functions $f$. Our suggestion is to consider smooth
functions and generalise the standard Lipschitz condition by requiring the
$L_2$-norm of the vector of all mixed derivatives of order $k\leq m$ to be
bounded above by a constant $L_m$, independent of the dimension of the domain
$\Omega$. 

Under such restrictions and an additional condition of independence of
predictor variables $x_1,x_2,\ldots,x_n$, 
we develop a technique for obtaining approximating
additive functions of a prescribed order and derive upper error bounds
in the $L_2$-norm (Section \ref{best}.) 
Our results are illustrated by a series of examples
in the last Section \ref{examples}, in particular
showing that the asymptotic rate of convergence
of the theoretically derived error is accurate.

Section \ref{quasi} aims at relaxing the assumption of 
independence of predictor variables. Recall that random variables
$x_1,x_2,\ldots,x_n$ are independent if the probability
distribution, $p(x)$, can be written as the product of the
marginal distributions, $p(x)=p_1(x_1)p_2(x_2)\ldots p_n(x_n)$. 
We replace this with the assumption which we call
{\it quasi-independence} and which calls for the distribution
of $x$ to be equivalent to the product of its marginals. The Radon-Nikodym
theorem then implies that the distribution
$\prod_{i=1}^n p_i(x_i)$ is the product of $p(x)$ with the
Radon--Nikodym derivative $\psi(x)$, and the additive approximation
obtained using the product measure (distribution) serves at the same
time as an approximation with regard to the `true' probability
distribution, $p(x)$. The upper error bounds
in the $L_2$-norm includes the derivative $\psi(x)$.

The research here is motivated by the observation that adaptive
techniques like MARS~\cite{Fri91} which estimate models of the form
given in Equation~(\ref{eq:anova}) will produce models with
predominantly lower order interactions. In practice, interactions with
order higher than 5 are not used. Models of the type defined in
Equation~(\ref{eq:anova}) have been called ``ANOVA decomposition''
in~\cite{EfrS81} as they generalise for real variables the models
which are used in the analysis of variance (ANOVA). Applications of
the ANOVA decomposition for the analysis of techniques for variance
estimation can be found in~\cite{EfrS81} and for the estimation of
quadrature errors in~\cite{Owe92} and~\cite{Hic96}. The work here
extends the previous work by providing estimates for the approximation
errors of truncated ANOVA decompositions. 
In earlier work by one of the authors~\cite{AndH99} it was seen 
how the concentration of measure may be exploited to get highly effective 
numerical differentiation procedures. Computational techniques for the 
determination of the ANOVA decomposition can for example be found 
in~\cite{Wah90,Fri91,HT}.

More generally, data mining \cite{BerL97,CHSVZ97,FayPS96} 
is being developed for the analysis of large data sets
which appeared in business and science due to the fact that both
data acquisition and data storage have become inexpensive because of
the availability of cheap transducers and data storage devices.
Typically, data mining applications lead to very large data 
sets of high dimension, and 
high-dimensional problems are intrinsically difficult as they are
affected by the \emph{curse of dimensionality}~\cite{Bel61,Heg}. Both
queries and the identification of predictive models are very
time-consuming. At the same time, it turns out that the effects of high
dimensions are not only bad and some 
may be successfully exploited to lead to highly
effective algorithms (cf. e.g. \cite{IM,AndH99}). 
In the ideal case, high-dimensional data is just
data which contains high amounts of information and these added
amounts of information should intuitively lead to better algorithms.

\section{Guiding observations\label{basic}}
\subsection{The paradox of increasing distances}
The first basic concept is that of an~\emph{object}
$\omega\in\Omega$. Examples include shoppers of a retailer, insurance
customers or variable stars. These objects have many properties, some of which
are observable. The array of observed properties is the \emph{feature
vector} $\xv$. We assume here that $\xv\in\R^n$ but arrays containing
other types of components and even of mixed types occur as well. In
order to distinguish objects we require some quantitative notion of
difference or similarity between objects. It seems reasonable to
assume that the Euclidean
distance between two feature vectors, given by
$$
   \sqrt{\sum_{i=1}^n (x_i-y_i)^2},
$$ 
provides information about the
difference of the underlying objects. However, this leads straight to
the first paradox of \emph{increasing distances}: the typical distance
between two objects grows as we add new features,
that is,
\emph{distance grows with the number of features} $n$. In other
words, the more
one knows about the objects, the more different they seem to appear and
ultimately, the difference may become infinite.  While the increased
difference seems reasonable, the unboundedness of the distance is not, as 
intuitively two objects are only ``different to a certain
point''. Fortunately, this paradoxical growth of distances may be
easily cured by either normalising the Euclidean distance 
or else by scaling the variables in such a way that, for example, the
average distance between two feature vectors is $1$. As an example,
consider the Euclidean cube $[0,1]^n$; it is easy to see that the
average distance between two randomly chosen vectors
$\xv,\yv\in [0,1]^n$ (the \emph{characteristic size}
of the cube \cite{Gr}) 
is $O(\sqrt n)$, and thus a natural way to
normalise the Euclidean distance is
$$
   d(\xv,\yv) : =\sqrt{ \frac{1}{n}\sum_{i=1}^n (x_i-y_i)^2}.
$$
While this paradox is seemingly simple, it is necessary to consider,
and it is important that the dissimilarity is normalised in the way
suggested so as to put things in the proper perspective.

The predictor functions $f$ we are interested in will estimate the
probability with which a certain statement about the object is true,
and thus the range of $f$ is, typically, the unit interval $[0,1]$. 
Moreover, it is reasonable to assume that the function assumes close
values for close values of parameters. Usually this condition is
expressed by requiring $f$ to be \emph{Lipschitz}, that is, to
satisfy
\[\vert f(\xv) - f(\yv)\vert \leq L\cdot d(\xv,\yv),\]
where $L>0$ is a positive number called the \emph{Lipschitz constant.}

We will assume that the data points are drawn from a domain $\Omega$
with respect to some probability distribution $P$.
Thus, we view the domain $\Omega =(\Omega,d,\A, P)$ as 
a probability space equipped with a distance
(an mm{\it -space}, cf. \cite{Gr}).

\subsection{Concentration of measure and approximation by constants}
The simplest class of additive functions is that of zeroth order of
interaction, $k=0$, in which case the approximating
functions $f_{\mathrm{add}}$
are simply constants. It turns out that even the approximation
by constants admits a substantial theory if the domain is high-dimensional.
Such approximation improves as dimension grows, which
observation is at the core of the 
{\it phenomenon of concentration of measure on high-dimensional structures.}
(See \cite{Gr, M2, Ta} and
numerous references therein.) 

Let $(\Omega_n)_{n=1}^\infty$ be an infinite family of metric spaces equipped
with probability measures. Intuitively, the spaces $\Omega_n$ should be
thought of as having
asymptotically growing dimension. Assume that the distances are so
normalised that the characteristic size of each
$\Omega_n$ is $O(1)$. 
Let $f$ be a Lipschitz function on some space $\Omega_n$, and assume
for simplicity that the corresponding Lipschitz constant $L=1$, that is,
\[\abs{f(x-y)}\leq d(x-y)\]
for all $x,y\in\Omega_n$. 
The concentration phenomenon refers to the observation that
for many `natural' families $(\Omega_n)$ as above,
the probability that $f(x)$ differs from its 
expected value by less than $\e>0$ is at least
\begin{equation}
1 - C_1\exp(-C_2\e^2n),
\label{conce}
\end{equation}
where $C_1,C_2>0$ are constants only depending on the family of
spaces $(\Omega_n)_{n=1}^\infty$ in question.
Intuitively, it means that a `nice' function 
on a space of high dimension `concentrates' near one value.

As an example, consider hyperspheres. The Euclidean $n$-sphere of
unit radius is defined as
\[{\mathbb S}^n:=\{\xv\in\R^{n+1}\colon \norm \xv = 1\}.\]
The constants for the family $({\Bbb S}^n)_{n=1}^\infty$ are
\[C_1=\sqrt{\frac\pi 8}, \hskip 0.5cm C_2=\frac 12.\]
Similar estimates with varying constants
hold for the hypercubes (remember that the distance has to be
appropriately normalised), the Euclidean spaces with the Gaussian measure,
the Hamming cubes, the groups of unitary matrices, and so forth.
({\it Loco citato.})

\begin{figure}
  \centerline{{\epsfig{file=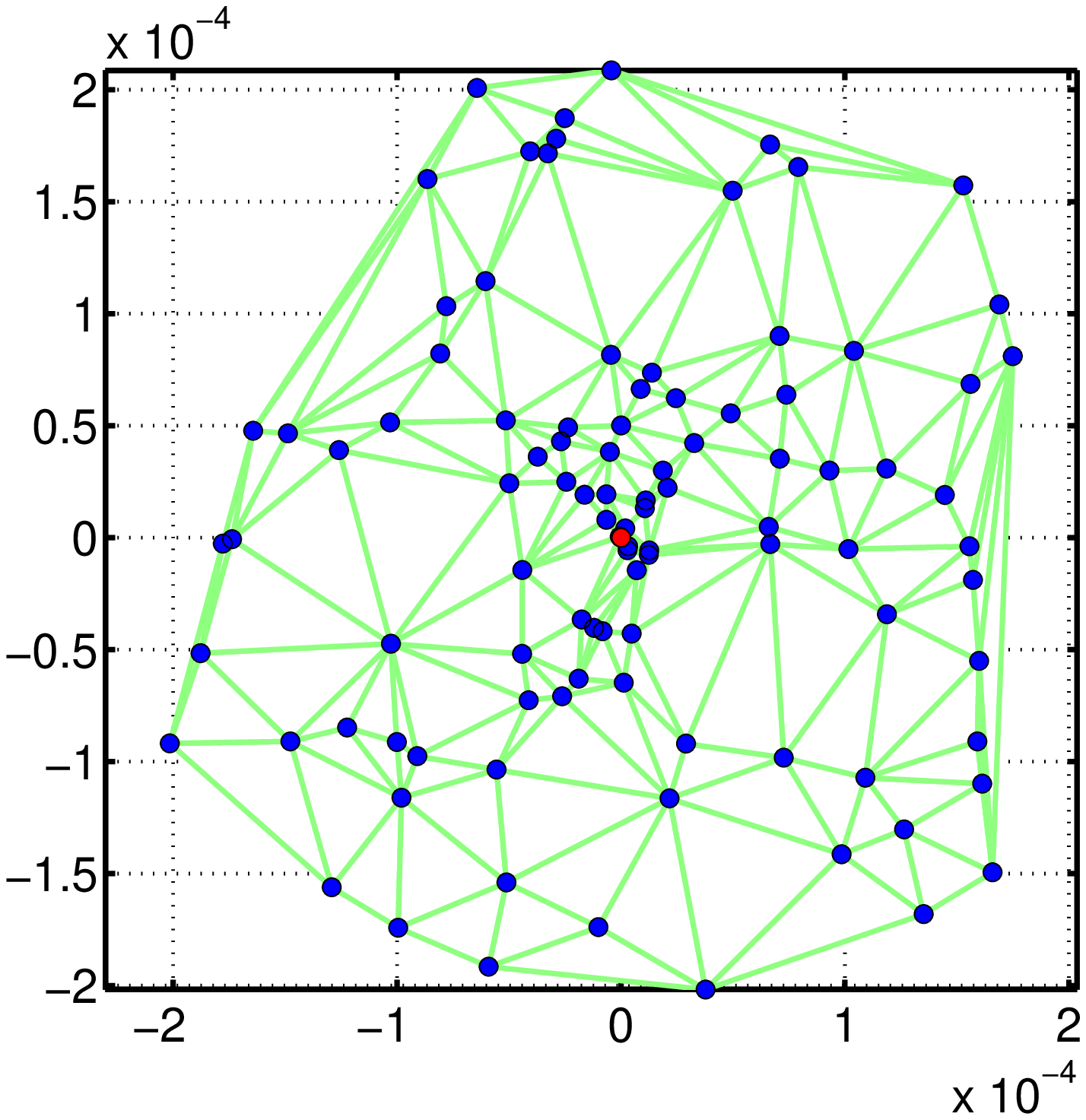,width=6cm}}
    {\epsfig{file=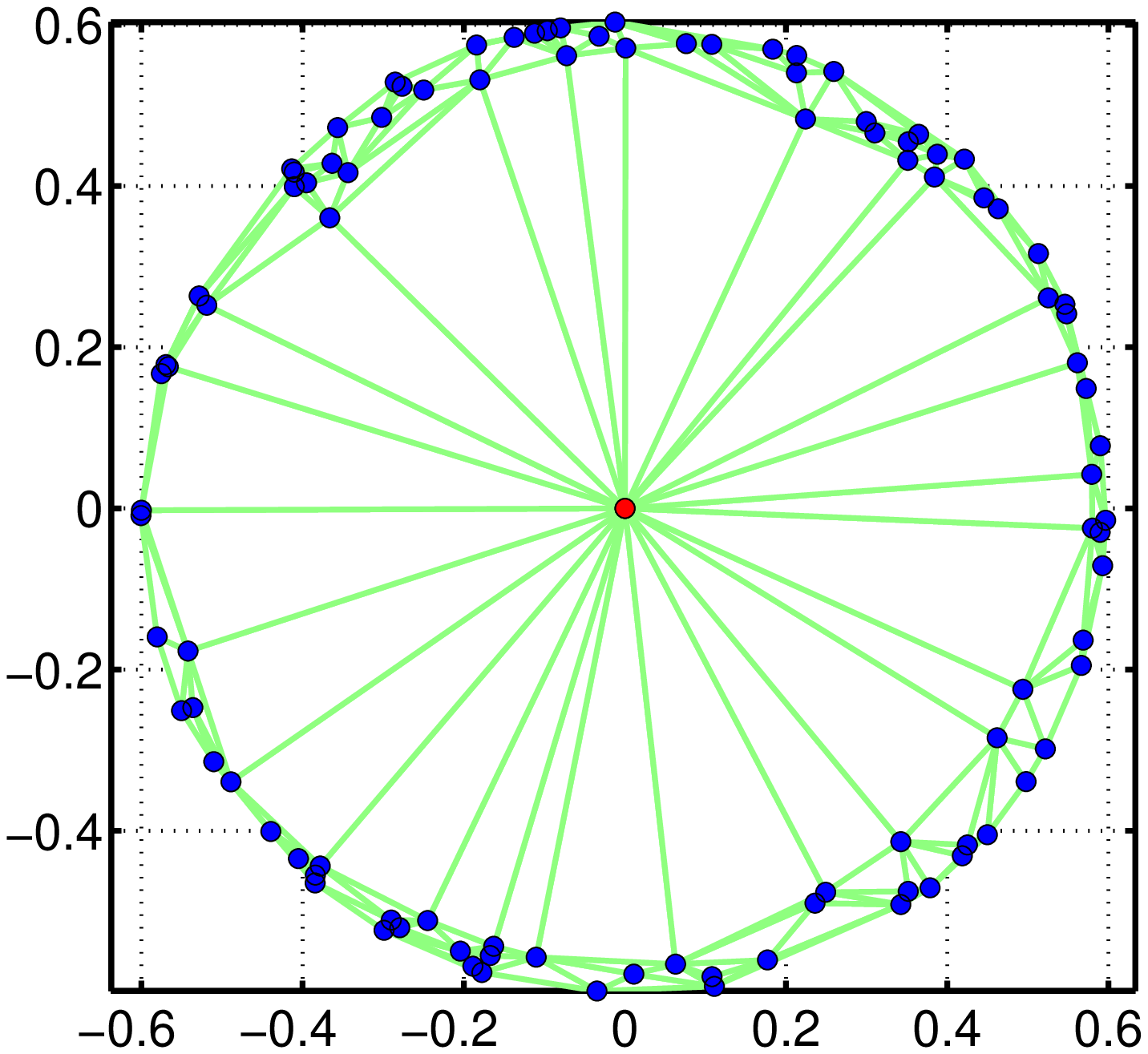,width=7cm}}}
   \caption{\label{fig:Conc} 100 nearest neighbors 
     of a point from i.i.d., normally
     distributed data points in 2 dimensions (left) and projected using
     a random distance preserving mapping from 100 dimensions (right).
     The edges are defined by a Delaunay triangulation.}
\end{figure}

Concentration of measure in particular 
enables one to explain the following, well-known,
observation made long ago about large and multidimensional datasets.
Consider an arbitrary object and its 
nearest neighbors. It can then be observed in practice and verified
theoretically that with higher dimensions \emph{the distances between the
point and its nearest neighbors become almost constant}, so that
the nearest neighbors all are close to a hypersphere around the
point. In fact, the large majority of the points seems to be contained
in a thin shell around the point. (See e.g. \cite{Bel61,BGR+99}.)
This paradox is illustrated in 
Figure~\ref{fig:Conc} where the hundred nearest neighbors of a point
are displayed all from an i.i.d. set of normally distributed data points.
To derive this paradox from the phenomenon of concentration of measure,
it is enough to apply the bound (\ref{conce}) to the function
$f(x)=d(x^\ast,x)$, where $x^\ast$ is the fixed (query) point. 
(See \cite{P5} for details.)
This paradox can \emph{not} be cured, at least not if one wants
to keep the same distance (similarity measure) between datapoints,
and is the origin of many problems
one faces with high-dimensional data. 

Returning back to the result captured by formula (\ref{conce}), one
concludes that the predictor function $f$ on a high-dimensional
domain will be closely approximated by a constant function, if the
dimension is large enough.
However, how good is such an approximation from
a practical point of view? It comes as no wonder that the dimension required
for the upper bound in (\ref{conce}) to become genuinely small is
extremely high. 

For example, suppose the data is uniformly distributed on the
hypersphere of dimension $n$ of unit radius,
$\Omega={\mathbb S}^n$, and the
predictor function $f\colon {\mathbb S}^n\to\R$ is $1$-Lipschitz and takes
values in the range $[-1,1]$. Let $\e>0$ and 
suppose we want to approximate $f$ by a 
constant, $M$, in such a way that $\abs{f(x)-M}<\e$ holds with probability
$>1-\e$. Even for the value of $\e=0.2$ (which is nowhere good enough)
the minimal dimension required to achieve our goal
is $n=41$. For $\e=0.1$, the minimal dimension
is $n=270$. To obtain the accuracy $\e=0.05$, the dimension $n=5000$
would suffice, but one can hardly expect a real dataset to have an
{\it intrinsic} dimension of this sort, that is, to depend on five
thousand independent parameters. Finally, to ensure that
\[P\{\abs{f(x)-M}<0.01\}>0.99,\]
which already seems to be a reasonably close approximation, one needs the
dimension to be on the order of the astronomical 
(and unrealistic) $n\sim 10^5$.

It is clear from the above that approximation by 
constants is not good enough
in the medium to high dimensions which is the case we are mostly interested in.
The next natural question is therefore: 
will the approximation error bounds
based on the concentration phenomenon and given by formula (\ref{conce})
improve automatically if one allows 
the approximation by additive functions
of {\it higher interaction order than zero}? 

It seems quite natural that by significantly 
relaxing the restrictions on the class of approximating functions one
gets better approximation bounds. Rather surprisingly, it is not the case,
as there exist functions on $n$-dimensional domains 
for which approximation by constants is the best
possible among {\it all} additive functions with interaction orders
$k$ of up to $n-1$. (Subsection \ref{zeroex}.)  

In view of the existence of such examples, it 
seems in a sense unavoidable that one should
impose additional restrictions on the predictor functions $f$ to
obtain better bounds on higher-order approximations with additive functions.
We will now put forward such restrictions as we find most natural, 
in the hope that the reader is prepared to accept them as such.

\subsection{Our assumptions}
Key to all approximation results are
assumptions about the data set and the class of functions to be
approximated. As we are interested in the asymptotic behaviour as the
dimension becomes medium to large (which means in practice larger than 10), we
actually are interested in a family of spaces, functions etc,
parameterised by their dimension. We will assume that the feature
vectors are distributed with a density $p(x)$ which has a first moment
$$
   E(x) = \int x p(x) dx
$$
and a variance
$$
E(\norm{x-E(x)}^2) = c
$$
which does not depend on the dimension. 

For some of the theorems it
will be required that the components of $x$ are independent, 
that is to say the underlying data distribution satisfies
$$
  p(x) = \prod_{i=1}^n p_i(x_i).
$$
However, for practical purposes this assumption is unrealistic, 
because in the context of data mining where one has many 
physical variables
($n$ is large), one would expect that those variables could be highly
correlated. In view of this,
we will subsequently replace the condition of independence
with a milder
restriction for the product distribution $p(x)$ to be {\it
in the same measure class} as the product distribution,
$$
  p(x) \sim \prod_{i=1}^n p_i(x_i).
$$
We will refer to such random variables $x_1,x_2,\ldots,x_n$ as
{\it quasi-independent.}

The functions we consider are assumed to be Lipschitz-continuous, i.e.,
$$
  |f(x) - f(y)| \leq L \|x-y\|
$$
for some constant $L$ which is independent of the dimension. In the case
of differentiable functions $f$ this corresponds to the bound
$$
   \sum_{i=1}^n \left(\frac{\partial f(x)}{\partial x_i}\right)^2 \leq L^2.
$$
This condition is very natural and invoked frequently, it assumes that
function values have similar sizes for points which are close. 

``Smoother'' functions will be defined as functions satisfying the condition
$$
  \sum_{1\leq i_i < \cdots < i_m \leq n} \left(\frac{\partial^m f(x)}%
    {\partial x_{i_1}\cdots \partial x_{i_m}}\right)^2 \leq L_m^2
$$
for some constant $L_m$ which is independent of the dimension $n$. While
such smoothness definitions do always have a certain degree of arbitrariness
and are difficult to check in applications, one can see, first, that this
condition generalises the Lipschitz condition, and, second, that for the case
of functions given as products
$$
  f(x) = \prod_{i=1}^n f_i(x_i)
$$
this bound follows from Lipschitz continuity for functions bounded away from
zero as one can verify that 
$$
   f(x)^{2(m-1)} 
  \sum_{1\leq i_i < \cdots < i_m \leq n} \left(\frac{\partial^m f(x)}%
    {\partial x_{i_1}\cdots \partial x_{i_m}}\right)^2 
   \leq 
\left(\sum_{i=1}^n \left(\frac{\partial f(x)}{\partial x_i}\right)^2 \right)^m.
$$
In particular, if $f(x) \geq 1 $ then $L_m = L^m$. Similar bounds for
the components of the decomposition from Equation~(\ref{eq:anova}) are used
in~\cite{Hic96} for the analysis of quadrature formulas based on a
reproducing kernel Hilbert space. We make here the small step to
consider families of functions where the bounds on the derivatives are
independent of the dimension in order to obtain an estimate of the
approximation error behaviour as a function of dimension.

So far we have assumed that the expected norm squared of the vectors
$x$ does not grow with dimension. However, in many practical
applications this is not the case. Often, the features are normalised,
say, so that they are all in $[-1, 1]$. This is the normalisation we will
mostly consider here if not mentioned otherwise. In this and many similar
cases the variance grows proportional to the dimension $n$.  Thus by
dividing all the variables by $\sqrt{n}$ one gets to the previous
situation again. If one invokes the Lipschitz condition in the
transformed variables, one has for the gradients in the original
variables the condition
\begin{equation}
  \label{eq:bound1}
  \sum_{i=1}^n 
    \left(\frac{\partial f}{\partial x_i}\right)^2 \leq \frac{L^2}{n}
\end{equation}
which is thus equivalent to the Lipschitz condition. For the higher order 
derivatives one gets the condition 
$$
  \sum_{1\leq i_i < \cdots < i_m \leq n} \left(\frac{\partial^m f(x)}%
    {\partial x_{i_1}\cdots \partial x_{i_m}}\right)^2 \leq 
    \frac{L_m^2}{n^m}.
$$
The dependence of the bounds on the dimension $n$ appears unnatural at
first, however note that this is simply a consequence of the scaling of
the variables, corresponding to Lipschitz-continuity in the unit ball.
The alternative to the introduction of this scaling in the conditions
would be to scale the variables so that the average distances would
be independent of the number of variables. For convenience, we have
here chosen to normalise all the $x_i$ to $[0,1]$ and to scale the
smoothness conditions.
While these conditions are strong, they are not unusual, and allow the
generalisation of the well-known limit theorems for 
$$
  f(x) = \frac{1}{n} \sum_{i=1}^n x_i
$$
of the average of $n$ i.i.d. random variables. A nontrivial example, for
which the conditions hold is given by 
$$
  f(x) = \exp\left(-\sum_{i=1}^n x_i^2 / n\right).
$$

A far-reaching but so-far implicit assumption is that all the
variables $x_i$ contribute in the same way to $f$. This is true for
applications where the features $x_i$ correspond to equally
important parts or elements and occur when $f$ is describing an
aggregate of similar elements each characterised by one $x_i$. Examples
include employees of a company and stars of a galaxy. An alternative
situation is considered in~\cite{NovW00,SloW98,Woz99} by H.~Wo\'zniakowski
and his collaborators. There the variables are
given weights depending on their importance. These weights then enter
the smoothness assumptions.  One of the consequences of that choice is
that in many cases the negative
effects of the concentration effect discussed
below can be avoided. Here, however, we consider a different situation
of equally important variables and thus have to deal with the 
consequences of the concentration.
The two examples of functions given above illustrate that functions 
like the mean which ``change'' with the dimension $n$ are very 
natural.

\section{Best approximations with additive functions\label{best}}
In this section we formulate the main results of this paper. 

\subsection{Independent random variables\label{independent}}
First the notation is established. 
 Let $(\Omega,\mathcal{A},P)$ be a
probability space, $x_1,\ldots,x_n$ denote a family of random
variables and $E(f\mid x_{i_1},\ldots,x_{i_k})$ be the usual
conditional expectations. Throughout this Subsection, 
we make a standing assumption that the random
  variables $x_1,x_2,\cdots,x_n$
are independent, i.e., the density distribution is of the
  form
\begin{equation}
  p(x) = \prod_{i=1}^n p_i(x_i)\mbox{ with } 
\int_{-\infty}^\infty p_i(x_i) dx_i = 1.
\label{density}
\end{equation}

The operator $D_i$ is defined as
$$
  (D_i f)(x) = f(x) - E(f\mid x_1,\ldots,x_{i-1},x_{i+1},\ldots,x_n).
$$
Using the independence assumption on random variables,
one gets a `telescoping sum'
\begin{equation}
  \label{eq:expansion1}
  f(x) = E(f) + \sum_{i=1}^n D_i E(f|x_1,\ldots,x_i).
\end{equation}
The terms of the sum can now be expanded in the same way as $f$ and
repeated application of these expansions provides a theorem which looks
very much like Taylor's theorem:
\begin{theorem}
\label{taylor}
  Let $f$ be an integrable function on $\Omega$. Then for every natural
$m$, $1\leq m\leq n$:
  \begin{equation*}
  \begin{split}
    f(x) & = E(f) + \sum_{i=1}^n D_i E(f\mid x_i)
  + \sum_{1\leq i_2 < i_1 \leq n} D_{i_2} D_{i_1} E(f\mid x_{i_2}, x_{i_1})
  + \cdots \\
  + &\sum_{1\leq i_{m-1}<\cdots<i_1\leq n}D_{i_{m-1}}\cdots D_{i_1} 
           E(f\mid x_{i_{m-1}}, \ldots, x_{i_1}) \\
  + &\sum_{1\leq i_m<\cdots<i_1\leq n}D_{i_m}\cdots D_{i_1} 
           E(f\mid x_1,x_2,\ldots,x_{i_m}, x_{i_{m-1}},
x_{i_{m-2}}, \ldots, x_{i_1}). \\
  \end{split}
  \end{equation*}
\end{theorem}
\begin{proof}
The proof uses induction. First, the case $m=1$ is  just
Equation~(\ref{eq:expansion1}).

If the equality holds for $m=k-1$ then the first $k-1$ terms are the same
as for $m=k$ and only the last term needs further expansion. In this term
each summand is a function of $x_{1},\ldots,x_{i_{k-1}-1}$ and 
from Equation~(\ref{eq:expansion1}) one obtains:
\begin{equation*}
\begin{split}
   & \sum_{1\leq i_{k-1}<\cdots<i_1\leq n}D_{i_{k-1}}\cdots D_{i_1} 
           E(f\mid x_1,\ldots,x_{i_{k-1}}, x_{i_{k-2}}, \ldots, x_{i_1}) \\
=   & \sum_{1\leq i_{k-1}<\cdots<i_1\leq n}D_{i_{k-1}}\cdots D_{i_1} 
           E(f\mid x_{i_{k-1}}, \ldots, x_{i_1}) \\
+   & \sum_{1\leq i_k<\cdots<i_1\leq n}D_{i_k}\cdots D_{i_1} 
           E(f\mid x_1,\ldots,x_{i_k}, x_{i_{k-1}}, \ldots, x_{i_1}).
\end{split}
\end{equation*}
Replacing the last term in the equation for the case $m=k-1$ with the
right-hand side of this equation leads to the equation for $m=k$.
\end{proof}

A similar decomposition for the special case of $m=n$ has been proved 
in~\cite{EfrS81} where the theorem is called \emph{Decomposition Lemma}. 

Next we introduce the space of $L_2$ functions which are sums of
functions only depending on $k$ variables each as:
$$
  L_{2,k} := \{ g(x) = \sum_{i_1,\ldots,i_k} 
       g_{i_1,\ldots,i_k}(x_{i_1},\ldots,x_{i_k})   \in L_2 \}.
$$
(Note that $L_{2,k}$ are closed, which follows from
Theorem \ref{projection} below.)
Now we introduce the operator $P_m : L_2 \rightarrow L_{2,m}$ such that
\begin{equation*}
  \begin{split}
  (P_m f)(x) =  
    E(f) & + \sum_{i=1}^n D_i E(f\mid x_i)
  + \sum_{1\leq i_2 < i_1 \leq n} D_{i_2} D_{i_1} E(f\mid x_{i_2}, x_{i_1})
  + \cdots \\
  & + \sum_{1\leq i_{m}<\cdots<i_1\leq n}D_{i_{m}}\cdots D_{i_1} 
           E(f\mid x_{i_{m}}, \ldots, x_{i_1}) 
\end{split}
\end{equation*}
and the remainder operator $R_m: L_2 \rightarrow L_{2,m}$ with
\begin{equation*}
  (R_{m} f) (x) =
   \sum_{1\leq i_{m}<\cdots<i_1\leq n}D_{i_{m}}\cdots D_{i_1} 
           E(f\mid x_1,\ldots, x_{i_{m}}, \ldots, x_{i_1}).
\end{equation*}
 From theorem \ref{taylor} one then gets $f=P_m f + R_{m+1}f$.

\begin{theorem}
\label{projection}
  The operator $P_m$ is an orthogonal projection,
and 
$$
E((f- P_mf)^2) \leq E((f-g)^2), 
\quad \text{for all $g\in L_{2,m}$
  and $f\in L_2$}.
$$
\end{theorem}

\begin{proof}
The statement that $P_m$ is a projection, 
i.e., $P_m^2 = P_m$, is equivalent to 
  showing that $R_{m+1}f$ is zero for $f\in L_{2,m}$. This follows
  directly from the fact that any function $f$ for which the function
  values do not depend on the variable $x_i$ one has $D_i f = 0$.
  As any $f\in L_{2,m}$ consists of a sum of functions which depend 
  only on $m$ variables and all the terms in $R_{m+1}$ contain 
  $D_{i_{m+1}}\cdots D_{i_1}$ there is at least one $i_k$ for which
  any particular term in the expansion of $f$ does not depend on
  $x_{i_k}$ and thus $R_{m+1}L_{2,m} = 0$.

If $p(x)$ is a product distribution as assumed above, then
  $\int p_i(x_i) dx_i = 1$ for every projection measure and for any
  $g(x_1,\ldots,x_n)$ one has
\begin{equation*}
  \begin{split}
  & \int D_i g (x_1,\ldots,x_n) p_i(x_i) dx_i  \\ = & 
  \int \left(g (x_1,\ldots,x_n)  -
  \int g(x_1,\ldots,x_{i-1},s,x_{i+1},\ldots,x_n) p_i(s)ds\right) 
    p_i(x_i) dx_i \\ = & 0.
  \end{split}
\end{equation*}
 
  Now as for each $m$-tuple $k_1<\cdots<k_m$ and for each $(m+1)$-tuple
  $i_{m+1}<\cdots<i_1$ one has at least one $i_j$ which is different
  from all the $k_s$ then 
\begin{equation*}
  \begin{split}
    & \int D_{i_{m+1}} \cdots D_{i_1} 
  E(f\mid  x_1,\ldots,x_{i_m}, x_{i_{m-1}}, \ldots, x_{i_1}) 
  g(x_{k_1},\ldots,x_{k_m}) p(x) dx \\
    = &\int \left( \int D_{i_{m+1}} \cdots D_{i_1} E(f\mid \cdots) 
     p_{i_j}(x_{i_j}) dx_{i_j}\right) g(x_{k_1},\ldots,x_{k_m})
     \prod_{t\neq i_j} p_t(x_t)dx_t \\
     = & 0.
  \end{split}
\end{equation*}
This shows that the error term $R_{m+1} f$ is orthogonal on $L_{2,m}$
which implies that $P_m$ is an orthogonal projection into $L_{2,m}$
and from this the minimisation characterisation follows.
\end{proof}

The orthogonality of all the components of the decomposition is shown
in~\cite{Owe92} for the case of the uniform distribution.

Now we observe that $R_n=0$ and thus the decomposition  as
in Theorem \ref{taylor} terminates and so
$$
f(x) = E(f) + \sum_{m=1}^n \;\;
  \sum_{1\leq i_{m}<\cdots<i_1\leq n}D_{i_{m}}\cdots D_{i_1} 
           E(f\mid x_{i_{m}}, \ldots, x_{i_1}).
$$
If all the variances of these terms exist, 
one has from the orthogonality of this decomposition
$$
\var(f) = \sum_{m=1}^n \;\;
  \sum_{1\leq i_{m}<\cdots<i_1\leq n}E\left(\left(\,D_{i_{m}}\cdots D_{i_1} 
           E(f\mid x_{i_{m}}, \ldots, x_{i_1})\,\right)^2\right).
$$

Error estimates are obtained for differentiable functions, one may also
get bounds based on Lipschitz constants. 
We introduce the (marginal)
cumulative distribution function
$$
    P_i(x_i) = \int_{-\infty}^x p_i(s) ds
$$
and the kernel
$$
  k_i(x_i,t_i) = P_i(t_i) - H(t_i-x_i)
$$
where $H(x)$ is the Heaviside function, i.e., $H(x)=1$ for $x\geq 1$ and
$H(x) = 0 $ for $x<0$. Using integration by parts one gets for 
differentiable $f$:
\begin{equation}
  \label{eq:repro}
  D_i f(x) = \int_{-\infty}^\infty k_i(x_i,t_i) 
     \frac{\partial f}{\partial t_i}(x_1,\ldots,x_{i-1},t_i,x_{i+1},
     \ldots,x_n)  dt_i.
\end{equation}
Now let $g_i(t_i)
:= \frac{\partial f}{\partial t_i}(x_1,\ldots,x_{i-1},t_i,x_{i+1},\ldots,x_n)
$ then the expected value squared is
$$
   E((D_if(x))^2) = \int k_i(x_i,t_i)k(x_i,s_i) g_i(t_i) g_i(s_i)
    p_i(x_i) dt_i ds_i dx_i.
$$
Now let
\begin{equation}
  G_i(t_1,t_2) := \min_{a,b=1,2} (P_i(t_a)(1-P_i(t_b))
\label{G}
\end{equation}
and
\begin{equation}
  \gamma := \int \max_i G_i(t,s) dt ds.
\label{gamma}
\end{equation}
Then one gets by integration by parts and from the Cauchy-Schwarz inequality:
$$
   \sum_{i=1}^n E((D_if(x))^2) \leq \gamma L^2
$$
if $f$ is Lipschitz continuous with constant $L$.

For the case of $m$ interactions we first define the seminorm $|f|_m$ by 
$$
  |f|_m^2 := \sup_{x} \sum_{1\leq i_1<\cdots< i_m \leq n} \left(
\frac{\partial^m f(x)}{\partial x_{i_1}\cdots\partial x_{i_m}}\right)^2
$$
and then obtain:
\begin{theorem}
Let $R_m$ be defined as in 
\begin{equation*}
  (R_{m} f) (x) =
   \sum_{1\leq i_{m}<\cdots<i_1\leq n}D_{i_{m}}\cdots D_{i_1} 
           E(f\mid x_1,\ldots, x_{i_{m}}, \ldots, x_{i_1}).
\end{equation*}
Then one has for the mean squared error bound
\begin{equation}
  E((R_{m} f)^2) \leq \gamma^m |f|_m^2.
\label{meanerror}
\end{equation}
\label{error}
\end{theorem}

\begin{proof}
  It is shown the same way as in an earlier theorem that 
  all the  terms of the sum defining $R_m f$ are orthogonal and so
$$
  E((R_mf)^2) = \sum_{1\leq i_1<\cdots< i_m \leq n}
  E\left( (D_{i_m}\cdots D_{i_1} 
  E(f \mid x_1,\ldots,x_{i_m},\ldots,x_{i_1}))^2\right).
$$
For simplicity set
$$
  g_{i_1,\ldots,i_m}(t_{i_1},\ldots,t_{i_m}) := 
 \frac{\partial^m  E(f \mid x_1,\ldots,x_{i_m-1},t_{i_m},%
     \ldots,t_{i_1})}      {\partial t_{i_1} \cdots \partial t_{i_m}}
$$
and application of Equation~(\ref{eq:repro}) and similar reasoning
as for the case $m=1$ gives
\begin{equation*}
  \begin{split}
  E((R_mf)^2)& = \\ = & \sum_{1\leq i_1<\cdots< i_m \leq n}
  \int   g_{i_1,\ldots,i_m}(t_{i_1},\ldots,t_{i_m})  
         g_{i_1,\ldots,i_m}(s_{i_1},\ldots,s_{i_m})
         \prod_{j=1}^m G_{i_j} (s_{i_j},t_{i_j})\, ds \, dt \\
    \leq & \;|f|_m^2\; \gamma_m
  \end{split}
\end{equation*}
where
$$
  \gamma_m := \int \max_{1\leq i_1<\cdots<i_m \leq n} \prod_{j=1}^m
    G_{i_j} (s_{i_j},t_{i_j}) ds dt.
$$
Now one can see that $\gamma_m \leq \gamma^m$ and from this the claimed
bound follows.
\end{proof}

Consider the case
$$
  p(x) = \prod_{i=1}^n q(x_i)
$$
for a fixed $q$ which are independent of $n$. For example, let the $x$
be uniformly distributed in the unit hypercube. Then the constant
$\gamma$ is independent of $n$. Furthermore, in 
Section \ref{basic} 
it was suggested that the appropriate smoothness restriction on $f$ is
$$
  |f|_m^2 \leq \frac{L_m^2}{n^m}.
$$
 From the previous theorem one can in this case conclude that
$$
  E((R_mf)^2) \leq \frac{ \gamma^m L_m^2}{n^m}.
$$
For example, if $q$ is the uniform distribution on $[-1,1]$ the constant
$\gamma$ can be computed explicitly and one gets $\gamma = \frac{1}{3}$
and so
$$
  E((R_mf)^2) \leq \frac{L_m^2}{3^m n^m}.
$$
In the case where $q$ is the normal distribution with expectation $0$
and variance $1$ one gets $\gamma = 0.516$ (rounded)
and thus 
$$
  E((R_mf)^2) \leq \frac{0.516^m L_m^2}{n^m}.
$$
The same bound is obtained if the components of $x$ are i.i.d. normal
with a variance such that $E(\norm x^2) = 1$ and $|f|_m \leq L_m$.

\subsection{\label{quasi} Quasi-independent random variables}
As we have already noticed, the assumption on independence of feature
vectors is not realistic for most practical data sets. Here we
propose one way to overcome this difficulty and get an approximation
which
\begin{enumerate}
  \item Satisfies a similar error bound as the one for the independent
        variable case.
  \item Has an error of the same order as the best least squares
        fit.
\end{enumerate}

 Let $\mu_1$ and $\mu_2$ be measures on the same sigma-algebra
of sets.
Recall that $\mu_1$ is {\it absolutely continuous} with regard to
$\mu_2$ if $\mu_2(A)=0$ always implies $\mu_1(A)=0$, that is,
every $\mu_2$-null set is also a $\mu_1$-null
set. The Radon--Nikodym theorem then implies that
$\mu_1 = \psi(x)\mu_2$, where $\psi(x)= d\mu_1/d\mu_2$ 
is a measurable function
called the {\it Radon--Nikodym derivative.}
The measures $\mu_1$ and
$\mu_2$ are {\it equivalent,} or {\it in the same measure class,} 
if they are absolutely continuous with respect to each other, 
that is, have the same null sets.

In our context, say that the random variables $x_1,x_2,\ldots,x_n$ 
on the probability space $(\Omega,\mathcal{A},P)$
are {\it quasi-independent} if the probability distribution $p(x)$ is
in the same measure class as the product of marginal distributions:

$$
  p(x) \sim \prod_{i=1}^n p_i(x_i).
$$

Denote by 
$P^\otimes$ the product measure on $X$ with the product distribution
above, and let
$$
\psi(x) =\frac{dP^\otimes}{dP}
$$
be the Radon--Nikodym derivative of the product
measure $P^\otimes$ with regard to the underlying measure 
$P$ on the data set. Then
one has
\begin{equation}
\prod_{i=1}^n p_i(x_i) = \psi(x) p(x).
\label{prodmeas}
\end{equation}

Given a predictor function $f$ on $X$, let $P^{\otimes}_m$ be the orthogonal
projection defined as in Subsection \ref{independent}, but using the
product measure $P^\otimes = \psi(x) P$ rather than the original 
probability distribution $P$. Then $P^{\otimes}_m(f)$ gives
an approximation of $f$ by a sum of additive functions of the order of
interaction $\leq m$. Denote
$$
R^{\otimes}_mf = f - P^{\otimes}_mf.
$$

Let $E^\otimes$ denote the expected value with regard to the
measure $P^\otimes$. Then for every random variable $y$ on $X$,
$$
E(y) = E^\otimes(y\cdot\psi^{-1}) 
\leq E^\otimes(y)\norm{\psi^{-1}}_\infty,
$$
where $\norm\cdot_\infty$ denotes the $L_\infty$-norm. 

Remember that the constant $\gamma$ only depends on the marginal
distributions $p_i(x)$, and so it remains the same no matter which
of the two measures, $P$ or $P^\otimes$, we are considering, cf.
(\ref{G}) and (\ref{gamma}). Now Theorem \ref{error} leads to
the following estimate.

\begin{theorem} The square error bound of the additive approximation 
satisfies
\begin{equation}
  E((R_{m}^\otimes f)^2) 
    \leq \gamma^m \norm{\frac{dP}{dP^\otimes}}_\infty |f|_m^2.
\label{r-nmeanerror}
\end{equation}
\end{theorem} 
\begin{proof}
Because of the equivalence of the measures the error in the original
measure is bounded by
$$
  E((R_{m}^\otimes f)^2) \leq \norm{\frac{dP}{dP^\otimes}}_\infty 
  E^\otimes((R_{m} f)^2)
$$
and Theorem~\ref{error} for the measure $P^\otimes$ then gives 
$  E^\otimes((R_{m} f)^2) \leq \gamma^m |f|_m^2$.
\end{proof}

Finally, the approximation does also provide a bound for the error 
of the best approximation in $L_{2,m}$:
\begin{theorem}
\begin{equation}
\min_{g\in L_{2,m}} E((f-g)^2) \leq E((R^\otimes_m f)^2) \leq \kappa 
\min_{g\in L_{2,m}} E((f-g)^2)
\end{equation}
where $\kappa =  \norm{\frac{dP}{dP^\otimes}}_\infty  
      \norm{\frac{dP^\otimes}{dP}}_\infty $.
\end{theorem}
\begin{proof}
The lower bound holds by definition. For the upper bound we first use
the property that the measures are in the same class to get:
$$
  E(R^\otimes_m f) \leq   \norm{\frac{dP}{dP^\otimes}}_\infty  
                         E^\otimes(R^\otimes_m f).
$$
Then we note that 
$P_m^\otimes$ is a best approximation with respect to the norm defined
by the product distribution by theorem~\ref{projection}. Thus one has
for any $g\in L_{2,m}$:
$$
  E^\otimes(R^\otimes_m f) \leq E^\otimes((g-f)^2)
$$
and, as the measures are in the same class:
$$
 E^\otimes((g-f)^2) \leq  \norm{\frac{dP^\otimes}{dP}}_\infty E((g-f)^2), 
$$
and combining these inequalities and taking the minimum over $g$ 
provides the desired estimate.
\end{proof}

This process does not lead directly to a computational procedure in the
case of dependent variables. We hope to
discuss such a procedure in a consequent paper.

\subsection{Extensions}

We have so far assumed that $f$ depends on exactly $n$ variables 
$x_1,\ldots,x_n$. Again in practice, any response variable $Y$ is 
typically only partially described by a function of the predictor
variables and a large proportion (in particular in data mining
applications) of $Y$ remains unexplained by $f(x_1,\ldots,x_n)$.
One way to model this situation is to assume that there are actually
$N>n$ predictor variables and by only limiting the approximations
to the first $n$ another error is introduced. The error of this
(truncation) approximation is
$$
   T_n f := f(x_1,\ldots,x_N) - E(f|x_1,\ldots,x_n).
$$
In this case the error of the approximation $P_{m,n}$ (previously
called $P_m$) has to include this term as well and thus the total
error is now
$$
  E_{m,n} f = R_{m,n} f + T_n f
$$
where $R_{m,n}$ (previously $R_m$) is the error term of the 
ANOVA decomposition of $E(f|x_1,\ldots,x_n)$.
These two error terms are orthogonal in the case of independent 
variables and one gets for the expected error squared:
$$
E((E_{m,n} f)^2)  = E((R_{m,n} f)^2) + E((T_n f)^2)
$$
Now the approximation in the space of functions of $n$ variables 
is the best possible and so is the additive approximation. Consequently,
by increasing $n$ the error is never increased. Clearly, if the error
of the proposed additive model is to be small then both the terms
$T_n f$ and $R_{m,n} f$ need to be small. In many practical cases,
however, the term $T_n f$ cannot be made small and thus major portions
of $f$ are beyond our controll. This, however, does not mean that the
approximation $P_{m,n}$ is of no practical use and being able to 
controll a portion of the variation of $f$ can lead to commercial
and scientific benefits.

Note that the smoothness condition for this case is really a smoothness
condition for $E(f|x_1,\ldots,x_n)$. An interesting question which
nevertheless has not been investigated here is how the smoothness of
the underlying function $f(x_1,\ldots,x_N)$ determines the smoothness
of the projection $E(f|x_1,\ldots,x_n)$.

\section{Examples \label{examples}}

In this section we will look at a few examples in more detail and we
will investigate how well the theory of the previous sections applies.
In the examples we will be considering approximations with constant
functions, with additive functions, and with functions with second
and third order interactions. The order of the interactions
is $m-1$ where $m$ has the values $1, 2, 3$ and $4$ respectively. 

\subsection{Example 1: uniform distribution on hypercube}
\begin{figure}
  \centerline{\epsfig{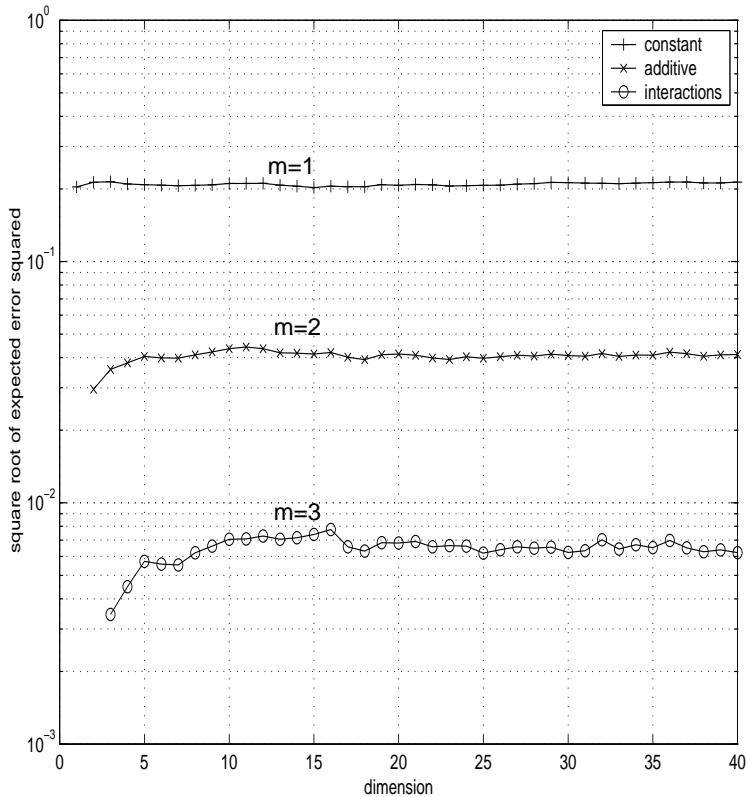}}
\caption{\label{fig:example1} $n^{m/2}$ times the square root of the average 
errors squared of a constant, additive and interaction approximation 
of $f(x) = \exp(-\sum_{i=1}^n x_i^2 /n)$.}
\end{figure}
For the uniform distribution on the hypercube $[-1,1]^n$ one has
$\gamma=1/3$. As a function to approximate we choose
$$
  f(x) = \exp\left(-\sum_{i=1}^n x_i^2 /n\right).
$$
 From this one gets
\begin{equation*}
 \begin{split}
  \sum_{1\leq i_1 < \cdots i_m \leq n} \left(\frac{\partial^m f(x)}%
    {\partial x_{i_1}\cdots \partial x_{i_m}} \right)^2
 = & \left(\frac{2}{n}\right)^{2m} f(x)^2
  \sum_{1\leq i_1 < \cdots i_m \leq n} x_{i_1}^2\cdots x_{i_m}^2 \\
  \leq & \left(\frac{2}{n}\right)^{2m} \binom{n}{m} \\
  \leq & \frac{4^m}{n^m m!}.
  \end{split}
\end{equation*}
Thus one can choose the Lipschitz constant to be
$$
  L_m^2 = \frac{4^m}{m!}
$$
and consequently the bound from the previous section is
$$
  E(R_m f^2) \leq \frac{4^m}{3^m m! n^m}.
$$
For practical error estimates this bound is slightly too pessimistic.
The order of convergence in $n$ is accurate. This can be seen from the
results of a simulation which are in Figure~\ref{fig:example1} where the
average errors squared have been multiplied by $n^{m/2}$ in order to confirm
the $O(n^{-m/2})$ behavior.

\subsection{Example 2: Function of three variables on data with bounded
 $E\left(\sum_i x_i^2\right)$}
\begin{figure}
  \centerline{\epsfig{file=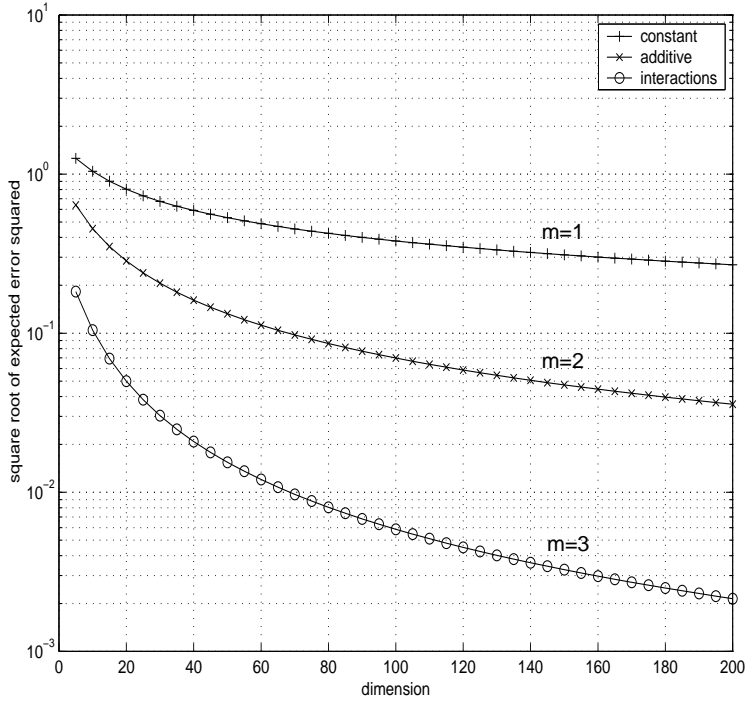,width=10cm}}
\caption{\label{fig:example2err} Square root of the average squared 
errors of a constant, additive and interaction approximation 
of $f(x) =  (1+\sin(x_1))(1+\sin(x_2))(1+\sin(x_3))$ for normally
distributed data with finite $E(\sum_i x_i^2)$.}
\end{figure}

\begin{figure}
  \centerline{\epsfig{file=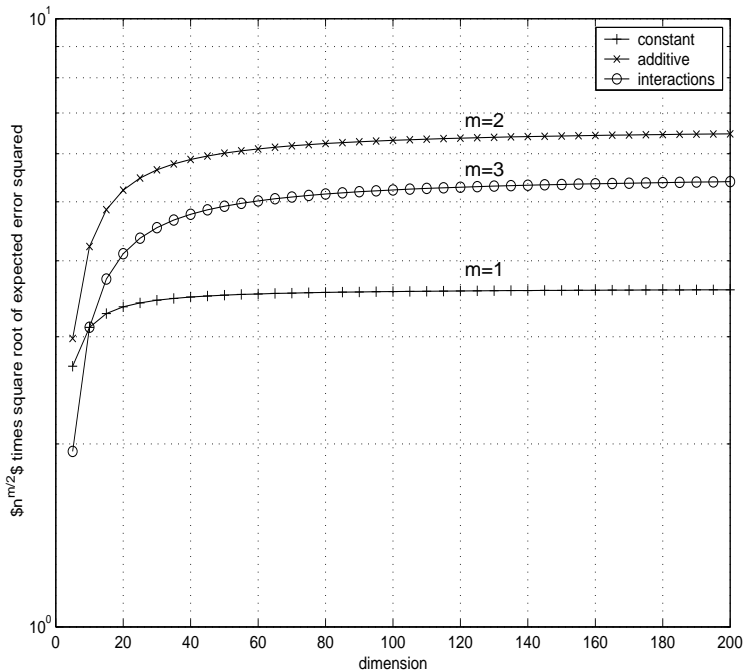,width=10cm}}
  \caption{\label{fig:example2} $n^{m/2}$ times square root of the average squared 
errors of a constant, additive and interaction approximation 
of $f(x) =  (1+\sin(x_1))(1+\sin(x_2))(1+\sin(x_3))$ for normally
distributed data with finite $E(\sum_i x_i^2)$.}
\end{figure}

Sometimes, functions are only dependent on a few of the variables. If
the data is equally distributed over many variables such that they 
have a uniformly (in the dimension) bounded expected squared norm
then the values of any component will concentrate around zero. This
is illustrated in this example. Here the function considered is
$$
  f(x) = (1+\sin(x_1))(1+\sin(x_2))(1+\sin(x_3)).
$$
The data points are assumed to be i.i.d. normally distributed and in
order to obtain the finite expected value of $\sum_i x_i^2$ the
variances of each component is $\sigma=1/n$. The error of the third
and higher order interaction approximations is zero, for the lower
order approximations see Figure~\ref{fig:example2err} for the
expected squared error. The theory again predicts asymptotic behaviour
of the error of $O(n^{-m})$ which is confirmed by the simulation result
displayed in Figure~\ref{fig:example2}.

\subsection{Example 3: Approximation obtained with MARS}

\begin{figure}
  \centerline{\epsfig{file=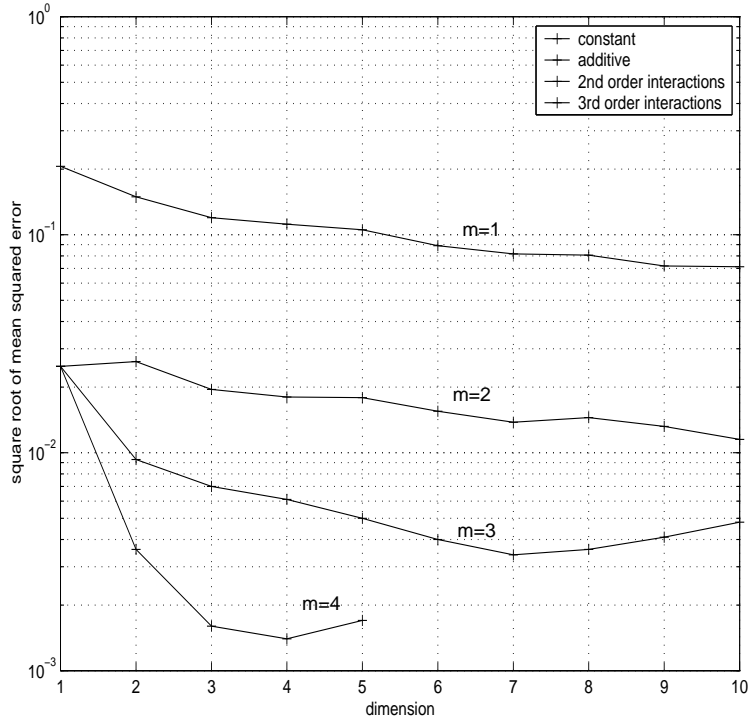,width=10cm}}
  \caption{\label{fig:mars1} Square root of the average squared 
errors of a constant, additive and interaction approximations
of $f(x) =  \exp(-\|x\|^2 /n)$ for data uniformly distributed on $
[-1,1]^n$. Approximation obtained with the MARS code.}
\end{figure}

The theory and the examples so far have illustrated the best possible
approximation with additive and interaction models. As in the first
example the function
$$
  f(x) = \exp\left({-\sum_{i=1}^n x_i^2 /n}\right)
$$
shall be approximated. We use 1000 data points uniformly distributed
on $[-1,1]^n$ and will let the dimension vary between 1 and 10. No 
random noise was added to $f(x)$. The approximations are computed with
the code ``MARS'' by J.Friedman~\cite{Fri91}. This code allows the
specification of the maximal order of interactions and we have
investigated the approximations obtained for constants, additive functions
and 2nd and 3rd order interaction models. The approximation uses tensor
products of piecewise linear functions and the basis functions used are
products of functions of the form
$$
  b(x_i) = (\pm (x_i - \xi))_+
$$
where $(x)_+ := \max(x,0)$. The total number of basis functions
allowed needs to be specified and we chose here $4\cdot n^m$ as an
upper limit such that basically each term in the ANOVA decomposition
may have $4$ basis functions on average. In Figure~\ref{fig:mars1} one
can see that allowing higher order interactions lead to better
approximations and also that in higher dimensions the approximations
have a tendency to get better with dimension. One can get an optimal
approximation if the number $m$ of interactions allowed equals the
dimension and thus one might expect a growth in error for dimensions
close to $m$. There is also a problem that allowing too many basis
functions of too high interaction may make the possible models too
complex and thus introduce instability for small enough data sizes.
The effect of allowing higher order interactions does not necessarily
mean that MARS will finally select terms with higher order
interactions at all. All these effects have to be taken into account
when one interprets Figure~\ref{fig:mars1}.

\subsection{\label{zeroex} Situations where constants give the best
additive approximation} 
The functions whose best additive approximation is by constants 
(and whose existence was claimed in Section \ref{basic})
are those possessing high degree of symmetry.
Without entering into technical details, let us mention two
examples which seem intuitively clear.

Let $\Omega=(\Omega,{\mathcal A},P)$ be a probability space which is
a domain in $\R^n$ and is such that the variables $x_1,x_2,\cdots,x_n$
are independent.

\begin{example}
Consider the function 
\[f(\xv)=x_1x_2\cdots x_n,\]
defined on the hypercube which is symmetric
about the origin, for example $\Omega=[-1,1]^n$. 
It can be proved that the best additive approximation in
the order of interaction $n-1$ is that by zero function.

Notice that in accordance with our philosophy the function $f$ has to be
normalised by the factor of $\frac 1n$, in order to keep its Lipschitz
constant bounded by $1$. The resulting function
\[f_1(\xv)=\frac 1 n x_1x_2\cdots x_n\]
on the same cube assumed pretty small values: its maximum is just
$\frac 1 n$, and thus one may argue that 
the approximation by zero function is not bad at all.
\end{example}

The next example is somewhat stronger.

\begin{example}
Denote by $\phi$ a usual bell-shaped function supported on the interval
$[-\frac 12 ,\frac 12 ]$, that is, a $C^\infty$ 
function taking values between $0$
and $1$, which is identically zero outside of $[-\frac 12 ,\frac 12 ]$, 
takes a positive value
at $0$, is monotone on each of the intervals $[-\frac 12 ,0]$ and
$[0,\frac 12 ]$, and satisfies $\phi^{(n)}(0)=0$ for all
natural $n$. Let us also assume that $\phi(0)=\frac 12$. 

For a vertex, $\e=(\e_1,\e_2,\cdots,\e_n)$, of the cube $[0,1]^n$ 
define the {\it parity} of $\e$
as the number of ones among the coordinates modulo $2$:
\[\abs\e := \sum_{i=1}^n\e_i \mod 2.\]
Now set for every $\xv\in [0,1]^n$
\[f(\xv) := \sum_{\e\in \{0,1\}^n} (-1)^{\abs\e}
\phi\left(\norm{\e-\xv}\right),\]
where $\norm\cdot$ denotes the Euclidean distance. A moment's thought
shows that $f$ is a well-defined $C^\infty$-function assuming
values in the interval between $-\frac 12$ and $\frac 12$, in particular
if $\xv=\e$ is a vertex, then
$f(\e)=\pm \frac 12$ depending on the parity.
Again, one can show that the above function admits no better additive
approximation in all orders of interaction up to $n-1$ inclusive
than that by the zero function.

 Notice that the normalisation of the function $f$ aimed at
keeping the Lipschitz constant of the order $O(1)$ 
leads to the function
\[f_1(\xv)=\frac 1 {\sqrt n} f(\xv),\]
whose maximal values reach $\frac 1 {2\sqrt n}$. 
\end{example}

\section{Conclusion}
In our paper we have attempted to perform initial analysis of the problem
of approximating a predictor function on a high-dimensional dataset with
additive functions allowing for interactions of a lower order. 
We are interested in the specifics of medium to high dimensions.
The proposed model makes what we believe to be reasonable assumptions,
from the modeling viewpoint, on the function to be approximated
(the normalisation conditions and 
`higher-order smoothness conditions'). We argue 
that some conditions of this kind are to be imposed in order
to obtain approximation results: we exhibit examples  
of Lipschitz functions in $n$ variables for which the best additive
function approximation of order of interaction $n-1$ is a constant.
Under the proposed conditions,
we derive from a Taylor-type theorem upper bounds on the approximation
errors. The results are illustrated on examples and compared to the
results obtained using the MARS software package. The examples confirm
that the asymptotic order of our error bounds is right. 

\section{Acknowledgements}
The second named author (V.P.) is grateful to the Computer Sciences Laboratory
of the Australian National University for hospitality extended 
between July and December 1999.
Part of the research (including the above visit)
was supported by the Australian Cooperative Research
Centre for Advanced Computational Systems (ACSys). 
Partial support was also provided 
by the Marsden Fund of the Royal Society of New Zealand,
in particular towards a 
visit by the first named author (M.H.) to
the Victoria University of Wellington in April 2002.
The authors acknowledge the constructive suggestions of the referee
which helped to substantially improve the paper.

\end{document}